\documentclass[12pt]{article}
\usepackage{graphicx}

\title{Quantum Spin Chain, Toeplitz Determinants and Fisher-Hartwig Conjecture}

\author{B.-Q.\ Jin and V.E.\ Korepin\\
{\small C.N.\ Yang Institute for Theoretical Physics}\\
 {\small State University of New York at Stony Brook}\\
 {\small Stony Brook, NY 11794-3840}}

\date{}
\begin{document}
\maketitle

\vspace{20mm}

\begin{abstract}
We consider one-dimensional quantum spin chain, which is called
\mbox{$\mathrm{XX}$} model (\mbox{$\mathrm{XX0}$} model or
isotropic \mbox{$\mathrm{XY}$} model) in a transverse magnetic
field. We study the model  on the infinite lattice at zero temperature.
We are  interested in the entropy of a subsystem
[a block of $\mathrm{L}$ neighboring spins] it describes entanglement
of the block with the rest of the ground state.
 For large blocks
entropy scales logarithmically \cite{vidal, vidal1}.
We prove the logarithmic formula for the leading term and
calculate the next term.
 We discovered that the
 dependence on the magnetic field interacting with spins is very simple:
the magnetic field effectively reduce the size of the subsystem.
 We also calculate entropy of a subsystem
of a small size. We also evaluated  R\'enyi and Tsallis  entropies of the
subsystem.  We represented the entropy in terms of a Toeplitz
determinant and calculated the asymptotic analytically.
\end{abstract}

\newpage
\section{Introduction}

In this paper we study entropy of a spin chain. The entropy is the
main object in thermodynamics and statistical physics. It is also
interesting for information theory \cite{ben,renyi,abe}. We study
the how entropy of a subsystem scales with the size of the
subsystem. We discover an interesting dependence on the magnetic
field interacting with spins. The physical system we consider is
\mbox{$\mathrm{XX}$} model in a transverse magnetic field.
 Hamiltonian for this model can be written as
\begin{eqnarray}
H_{\mathrm{XX}}(h)=- \sum_{n=1}^N (
\sigma^x_{n}\sigma^x_{n+1}+\sigma^y_{n}\sigma^y_{n+1} +
h\sigma^z_{n}). \label{xxh}
\end{eqnarray}
Here $\sigma^x_n$, $\sigma^y_n$, $\sigma^z_n$ are Pauli matrix,
which describe spin operators on $n$-th lattice site, $h$ is the
magnetic field and $N$ is the number of total lattice sites of
spin chain (we take $N\to \infty$ in this paper). This model has
been solved by E.\ Lieb, T.\ Schultz and D.\ Mattis in
zero-magnetic field case \cite{Lieb} and by E.\ Barouch and B.M.\
McCoy in the presence of a constant magnetic field \cite{mccoy}.
Some exact calculation of time-dependent properties also exists,
for examples, see Ref.~\cite{mccoy2} by E.\ Barouch, B.M.\ McCoy
and Refs.~\cite{gallavotti} by M.\ Dresden and D.B.\ Abraham, E.\
Barouch, G.\ Gallavotti and A.\ Martin-L\"of. We shall consider the model in
 thermodynamic limit $N\rightarrow \infty$.
The ground state is ferromagnetic for $|h|> 2$ while it's critical
for $|h|<2$. The ground state $|GS\rangle$ is unique. So the
entropy of the whole infinite ground state is zero, but it can be
positive for a subsystem [a part of the ground state]. We shall
calculate the entropy of  $\mathrm{L}$ neighboring spins. We shall
call the first $\mathrm{L}$ neighboring spins as sub-system A and
the rest as sub-system B. We shall consider von
 Neumann entropy
($S(\rho_A)$),  R\'enyi \cite{renyi} and Tsallis \cite{tsallis} entropies
 for sub-system A:
\begin{eqnarray}
S_{\mbox{von Neumann }}=S(\rho_A)&=&-Tr(\rho_A \ln \rho_A), \label{edif}\\
S_{\mbox{R\'enyi}}=S_{\alpha}(\rho_A)&=&\frac{1}{1-\alpha}\ln Tr(\rho_A^{\alpha}),
\qquad \alpha\neq 1 \quad\textrm{and}\quad \alpha>0 .\label{olds}\\
S_{\mbox{Tsallis}}&=& {{\mbox{Tr}}\rho^\alpha -1\over 1-\alpha}\label{tsa}
\end{eqnarray}
Von Neumann entropy is the standard one. Tsallis  and R\'enyi entropy may
also be important for both information theory and statistical
physics, see  \cite{tsallis} and \cite{renyi}. When $\alpha \to 1$, Tsallis and  R\'enyi
entropy turns into von Neumann entropy.\footnote{Tsallis entropy
and   R\'enyi entropy are algebraically related: $S_{\mbox{Tsallis}}= {e^{(1-\alpha)S_\alpha}-1 \over 1-\alpha}$
 } Here \mbox{$\rho_A$}
 is the
reduced density matrix of sub-system A:
$$\rho_A=Tr_B(\rho_{AB})$$
 and the density matrix of the whole system is
$$\rho_{AB}=|GS\rangle \langle GS|$$
 for zero temperature.
Since calculations for von Neumann entropy and R\'enyi entropy are
much similar, we give the detail calculation for von Neumann
entropy only. The explicit result for R\'enyi entropy will be
given without derivation.

G. Vidal, J.I. Latorre, E. Rico, and A. Kitaev emphasized the role of
the entropy of the subsystem in information theory [it describes entanglement of the subsystem with the rest of the ground state]. They showed that
for subsystems of the large size $L$ von Neumann entropy scales logarithmically
$S(\rho_A) \sim (1/3)\ln L$, see \cite{vidal, vidal1}. In this paper we prove
 this formula and 
calculate the next term of asymptotic decomposition. We also evaluated
 R\'enyi entropy of the subsystem.
Before we give the full derivation in the following sections, we
 first summarize our results here. We discovered  that
one can introduce a universal scaling variable:
$$\mathcal{L}=2\mathrm{L}\sqrt{1-\left(\frac{h}{2}\right)^2}$$
We consider the magnetic field to be less then critical value for
$|h|<h_c$. When  the magnetic field is larger then the critical
value $h_c=2$ then the ground state is ferromagnetic [all spins
are parallel], it does not have any entropy. We calculated von
Neumann entropy and R\'enyi entropy of block spins for   the
magnetic field smaller then critical value. Here we present the
expression for entropies for large $\mathcal{L}$ and small
$\mathcal{L}$
\begin{eqnarray}
S_{\alpha}(\rho_A) \approx \left\{ \begin{array}{ll} \left\{
\begin{array}{ll} \frac{1}{1-\alpha} \ln
 \left(\left( \frac{\mathcal{L}}{2 \pi}\right)^{\alpha}+
 \left(1- \frac{\mathcal{L}}{2 \pi}\right)^{\alpha}\right) & \textrm{($\alpha\ne 1$)}\\
\frac{\mathcal{L}}{\pi} \ln \frac{\pi}{\mathcal{L} } & \textrm{($\alpha=1$)}\\
\end{array} \right. & \textrm{if $0<\mathcal{L}
< 1$}\\
\frac{1+\alpha^{-1}}{6} \ln \mathcal{L} +\Upsilon_1^{\{\alpha\}} &
\textrm{if $\mathcal{L}\gg 1$}
\end{array} \right.
\end{eqnarray}
Here $\Upsilon_1^{\{\alpha\}}$ is a constant defined in
Eq.~\ref{up11}. When $\alpha=1$, R\'enyi entropy
$S_{\alpha}(\rho_A)$ becomes von Neumann entropy, the coefficient
for $\log \mathcal{L}$ in large $\mathcal{L}$ expression becomes
$\frac{1}{3}$ and $\Upsilon_1^{\{\alpha\}}$ becomes
\begin{eqnarray}
 \Upsilon_1& = & - \int_0^\infty \mathrm{d} t \left\{ {e^{-t} \over 3 t}
  + {1 \over  t \sinh^2 (t/2)} - { \cosh (t/2) \over 2 \sinh^3 (t/2)}
  \right\}.\nonumber
\end{eqnarray}

Following Ref.~\cite{Lieb}, let us introduce two Majorana
operators
\begin{eqnarray}
c_{2l-1}= (\prod_{n=1}^{l-1} \sigma^z_{n}) \sigma^x_l~~~\textrm{and}
~~~c_{2l}= (\prod_{n=1}^{l-1} \sigma^z_{n}) \sigma^y_l,
\end{eqnarray}
on each site of the spin chain.
Operators $c_n$ are hermitian and obey the anti-commutation relations \mbox{$\{ c_m, c_n\} =2 \delta_{mn}$}. In terms of
operators $c_n$, Hamiltonian $H_{\mathrm{XX}}$ can be rewritten as
\begin{eqnarray}
H_{\mathrm{XX}}(h)= i \sum_{n=1}^N (c_{2n}c_{2n+1}- c_{2n-1}c_{2n+2}+ h c_{2n-1}c_{2n}).
\end{eqnarray}
Here different boundary effects can be ignored because we are only interested in cases with \mbox{$N\to \infty$}.
This Hamiltonian can be subsequently diagonalized by linearly transforming the operators $c_n$. It has been
obtained \cite{Lieb,mccoy} (also see \cite{vidal,vidal1}) that
\begin{eqnarray}
\langle GS| c_m|GS\rangle=0,~~\langle GS|c_m c_n|GS\rangle =\delta_{mn}+i (\mathbf{B}_N)_{mn}.
\end{eqnarray}
Here matrix $\mathbf{B}_N$ can be written in a block form as
\begin{eqnarray}
\mathbf{B}_N=\left( \begin{array}{cccc}
\Pi_0 &\Pi_{-1}& \ldots &\Pi_{1-N}\\
\Pi_{1}& \Pi_0&   &   \vdots\\
\vdots &      & \ddots&\vdots\\
\Pi_{N-1}& \ldots& \ldots& \Pi_0
\end{array}     \right) \quad\textrm{and}\quad
\Pi_l=\frac{1}{2\pi} \int_{0}^{2\pi} \, \mathrm{d} \theta\, e^{-\mathrm{i} l \theta} {\cal G}(\theta) \label{bn},
\end{eqnarray}
where both $\Pi_l$ and ${\cal G}(\theta)$ (for \mbox{$N \to
\infty$}) are \mbox{$2\times 2$} matrix,
\begin{eqnarray}
{\cal G}(\theta)=\left( \begin{array}{cc}
               0& g(\theta)\\
               -g(\theta)&0
               \end{array} \right),~~g(\theta)=\left\{ \begin{array}{rl}
1, & -k_F < \theta <k_F,\\
-1, & k_F < \theta < (2\pi-k_F)
\end{array} \right.
\end{eqnarray}
and $k_F=\arccos(|h|/2)$. Other correlations such as
\mbox{$\langle GS|c_m \cdots c_n|GS\rangle$} are obtainable by
Wick theorem. The Hilbert space of sub-system A can be spanned by
\mbox{$\prod_{i=1}^{\mathrm{L}}
\{\sigma^{-}_i\}^{p_i}|0\rangle_F$}, where $\sigma^{\pm}_i$ is
Pauli matrix, $p_i$ takes value $0$ or $1$, and vector
\mbox{$|0\rangle_F$} denotes the ferromagnetic state with all
spins up. We are also able to construct a set of fermionic
operators $b_i$ and $b^{+}_i$ by defining
\begin{eqnarray}
d_m= \sum_{n=1}^{2\mathrm{L}} v_{mn} c_n,~~m=1,\cdots, 2\mathrm{L};~~~
b_l= (d_{2l} +i d_{2l+1})/2, ~~l=1,\cdots, \mathrm{L}\label{ctob}
\end{eqnarray}
with \mbox{$v_{mn}\equiv (\mathbf{V})_{mn}$}. Here matrix $\mathbf{V}$ is an orthogonal matrix.
It's easy to verify that $d_m$ is hermitian operator and
\begin{eqnarray}
b^+_l= (d_{2l} -i d_{2l+1})/2,~~~\{b_i,b_j\}=0,~~~\{b^{+}_i,b^{+}_j\}=0,~~~\{b^{+}_i,b_j\}=\delta_{i,j}.
\end{eqnarray}
In terms of fermionic operators $b_i$ and $b^{+}_i$, the Hilbert space can also be spanned
by \mbox{$\prod_{i=1}^{\mathrm{L}} \{b^{+}_i\}^{p_i}|0 \rangle_{vac}$}. Here $p_i$ takes value $0$ or $1$,
$2\mathrm{L}$ fermionic operators $b_i$, $b^{+}_i$ and vacuum state \mbox{$|0\rangle_{vac}$} can be constructed by requiring
\begin{eqnarray}
b_l|0\rangle_{vac}=0,~~l=1,\cdots, \mathrm{L}.
\end{eqnarray}
We shall choose a specific orthogonal matrix $\mathbf{V}$ later.

\section{Density Matrix of Sub-system A}

Let $\{\psi_I\}$ be a set of orthogonal basis for Hilbert space of
any physical system. Then the most general form for density matrix
of this physical system can be written as
\begin{eqnarray}
\rho=\sum_{I,J} c(I,J) |\psi_I\rangle \langle\psi_J|.
\end{eqnarray}
Here \mbox{$c(I,J)$} are complex coefficients. We can introduce a
set of operators \mbox{$P(I,J)$} by $ P(I,J) \propto
|\psi_I\rangle \langle\psi_J|$ and \mbox{$\widetilde{P}(I,J)$}
satisfying
\begin{eqnarray}
\widetilde{P}(I,J) P(J,K)=\delta_{I,K} |\psi_I\rangle \langle\psi_I|,~~ P(I,J) \widetilde{P}(J,K)=\delta_{I,K} |\psi_I\rangle \langle\psi_I|.
\end{eqnarray}
There is no summation over repeated index in these formula. We shall use an explicit summation symbol through the whole paper.
Then we can write the density matrix as
\begin{eqnarray}
\rho=\sum_{I,J} \tilde{c}(I,J) P(I,J),~~ \tilde{c}(I,J)=Tr(\rho \widetilde{P}(J,I)).
\end{eqnarray}
Now let us consider quantum spin chain defined in Eq.~$\ref{xxh}$.
For the sub-system A, the complete set of operators
\mbox{$P(I,J)$} can be generated by
\mbox{$\prod_{i=1}^{\mathrm{L}} O_i$}. Here we take operator $O_i$
to be any one of the four operators \mbox{$\{ b^{+}_i, b_i,
b^{+}_i b_i, b_i\, b^{+}_i\}$} (Remember that $b_i$ and $b_i^+$
are fermionic operators defined in Eq.~\ref{ctob}). It's easy to
find that \mbox{$\widetilde{P}(J,I)=(\prod_{i=1}^{\mathrm{L}}
O_i)^{\dagger}$}  if \mbox{$P(I,J)=\prod_{i=1}^{\mathrm{L}} O_i$}.
Here ${\dagger}$ means hermitian conjugation. Therefore, the
reduced density matrix for sub-system A can be represented as
\begin{eqnarray}
\rho_A=\sum Tr_{AB} \left(\rho_{AB} (\prod_{i=1}^{\mathrm{L}}
O_i)^{\dagger}\right) \prod_{i=1}^{\mathrm{L}} O_i.
\end{eqnarray}
Here the summation is over all possible different terms
\mbox{$\prod_{i=1}^{\mathrm{L}} O_i$}. For the whole system to be
in pure state \mbox{$|GS\rangle$}, the density matrix $\rho_{AB}$
is represented by \mbox{$|GS\rangle \langle GS|$}. Then we have
the expression for $\rho_A$ as following
\begin{eqnarray}
\rho_A=\sum \langle GS|(\prod_{i=1}^{\mathrm{L}} O_i)^{\dagger}|GS \rangle \prod_{i=1}^{\mathrm{L}} O_i\;. \label{ll19}
\end{eqnarray}
This is the expression of density matrix with the coefficients related to multi-point correlation functions. These correlation functions are
well studied in the physics literature \cite{korepin}.
Now let us choose matrix $\mathbf{V}$ in Eq.~$\ref{ctob}$ so that the set of fermionic basis $\{ b^+_i\}$ and $\{ b_i\}$ satisfy an equation
\begin{eqnarray}
\langle GS| b_i b_j|GS \rangle= 0,~~
\langle GS| b^+_i b_j|GS \rangle= \delta_{i,j} \langle GS|b^+_i b_i|GS\rangle. \label{ddb}
\end{eqnarray}
Then the reduced density matrix $\rho_A$ represented as sum of products in Eq.~$\ref{ll19}$ can be represented as a product of sums
\begin{eqnarray}
\rho_A= \prod_{i=1}^{\mathrm{L}} \Bigl( \langle GS|b^+_i b_i|GS
\rangle b^+_i b_i+\langle GS|b_i b^+_i|GS \rangle b_i b^+_i
\Bigr).\label{dmf}
\end{eqnarray}
Here we used the equations \mbox{$\langle GS| b_i|GS\rangle=0=\langle GS| b^+_i|GS\rangle$} and Wick theorem.
This fermionic basis was suggested by G.\ Vidal, J.I.\ Latorre, E.\
Rico and A.\ Kitaev in Ref.~\cite{vidal,vidal1}. A similar result for the
density matrix of a subsystem in terms of free spinless fermion
model was obtained by C.A.\ Cheong and C.L.\ Henley in Ref.~\cite{Henley}.

\section{Closed Form for The Entropy}

Now let us find a matrix $\mathbf{V}$ in Eq.~$\ref{ctob}$, which
will block-diagonalize correlation functions of Majorana operators
$c_n$. From Eqs.~$\ref{ctob}$ and $\ref{bn}$, we have the
following expression for correlation function of $d_n$ operators:
\begin{eqnarray}
\langle GS|d_m d_n|GS\rangle&=&\sum_{i=1}^{2\mathrm{L}} \sum_{j=1}^{2\mathrm{L}} v_{mi} \langle GS|c_i c_j|GS\rangle v_{jn} \;,\nonumber\\
\langle GS|c_m c_n|GS\rangle&=&\delta_{mn} +\mathrm{i} (\mathbf{B}_{\mathrm{L}})_{mn},\nonumber\\
\langle GS|d_md_n|GS\rangle&=&\delta_{mn}+ \mathrm{i} (\widetilde{\mathbf{B}}_{\mathrm{L}})_{mn}.\label{diaB}
\end{eqnarray}
The last equation is the definition of a matrix
$\widetilde{\mathbf{B}}_{\mathrm{L}}$. Matrix
$\mathbf{B}_{\mathrm{L}}$ is the sub-matrix of
$\mathbf{B}_{\mathrm{N}}$ defined in Eq.~\ref{bn} with
$m,n=1,2,\dots,\mathrm{L}$. We also require
$\widetilde{\mathbf{B}}_{\mathrm{L}}$ to be the form
\cite{vidal,vidal1}
\begin{eqnarray}
\widetilde{\mathbf{B}}_{\mathrm{L}}=V \mathbf{B}_{\mathrm{L}} V^T=
\oplus_{m=1}^{\mathrm{L}} \nu_m \left( \begin{array}{cc}
               0& 1\\
               -1&0
               \end{array} \right)= \mathbf{\Omega}\otimes \left( \begin{array}{cc}
               0& 1\\
               -1&0
               \end{array} \right).\label{mmf}
\end{eqnarray}
Here matrix $\mathbf{\Omega}$ is a diagonal matrix with elements
$\nu_m$ (all $\nu_m$ are real numbers). Therefore, choosing matrix
$\mathbf{V}$ satisfying Eq.~$\ref{mmf}$ in Eq.~$\ref{ctob}$, we
obtain $2\mathrm{L}$ operators $\{ b_l\}$ and $\{ b^+_l\}$ with
following expectation values
\begin{eqnarray}
\langle GS|b_m|GS \rangle=0\;,\langle GS|b_m b_n|GS\rangle =0\;,\langle GS|b^+_m b_n|GS \rangle =\delta_{mn}\frac{1+\nu_m}{2}.
\end{eqnarray}
Using the simple expression for reduced density matrix $\rho_A$ in
Eq.~$\ref{dmf}$, we obtain
\begin{eqnarray}
\rho_A= \prod_{i=1}^{\mathrm{L}} \left( \frac{1+\nu_i}{2} b^+_i b_i+ \frac{1-\nu_i}{2} b_i b^+_i \right).
\end{eqnarray}
This form immediately gives us all the eigenvalues $\lambda_{x_1x_2\cdots x_{\mathrm{L}}}$ of reduced density matrix $\rho_A$,
\begin{eqnarray}
\lambda_{x_1x_2\cdots x_{\mathrm{L}}}=\prod_{i=1}^{\mathrm{L}}
\frac{1+(-1)^{x_i} \nu_i}{2},~~~x_i=0,1~~\forall i. \label{eigf}
\end{eqnarray}
Note that in total we have $2^{\mathrm{L}}$ eigenvalues. Hence,
the entropy of $\rho_A$ from Eq.~$\ref{edif}$ becomes
\begin{eqnarray}
S(\rho_A)&=&\sum_{m=1}^{\mathrm{L}} e(1,\nu_m) \label{eaap}
\end{eqnarray}
with
\begin{eqnarray}
 e(x, \nu)= -\frac{x+\nu}{2} \ln (\frac{x+\nu}{2})-\frac{x-\nu}{2} \ln (\frac{x-\nu}{2}).\label{intee}
\end{eqnarray}
We shall use this result further to obtain analytical asymptotic.
Function \mbox{$e(1, \nu)$} in Eq.~$\ref{eaap}$ is equal to the
Shannon entropy function \mbox{$H(\frac{1+\nu}{2})$}. However, in
the following calculation (Eq.~$\ref{eaa}$), we will need the more
general function \mbox{$e(x, \nu)$} instead of \mbox{$H(\nu)$}.
Notice further that matrix $\mathbf{B}_{\mathrm{L}}$ can have a
direct product form, i.e.
\begin{eqnarray}
\mathbf{B}_{\mathrm{L}}= \mathbf{G}_{\mathrm{L}} \otimes \left(
\begin{array}{cc}
0& 1\\
-1&0
\end{array} \right)\quad \label{bgm}\textrm{with} \quad\mathbf{G}_{\mathrm{L}}=\left( \begin{array}{cccc}
g_0 &g_{-1}& \ldots &g_{1-L}\\
g_{1}& g_0&   &   \vdots\\
\vdots &      & \ddots&\vdots\\
g_{L-1}& \ldots& \ldots& g_0
\end{array}     \right)\;, \label{mg1}
\end{eqnarray}
where $g_l$ is defined as
\begin{eqnarray}
g_l=\frac{1}{2\pi} \int_0^{2\pi} \, \mathrm{d} \theta\,
e^{-\mathrm{i} l \theta} g(\theta) ~~\textrm{and}~~
g(\theta)=\left\{ \begin{array}{rl}
1, & -k_F < \theta <k_F,\\
-1, & k_F < \theta < (2\pi-k_F)
\end{array} \right.\label{mg2}
\end{eqnarray}
and \mbox{$k_F=\arccos(|h|/2)$}. From Eqs.~$\ref{mmf}$ and
$\ref{bgm}$, we conclude that all $\nu_m$ are just the eigenvalues
of real symmetric matrix $\mathbf{G}_{\mathrm{L}}$.

\begin{figure}[ht]
\begin{center}
\includegraphics[width=360pt,height=!,angle=0]{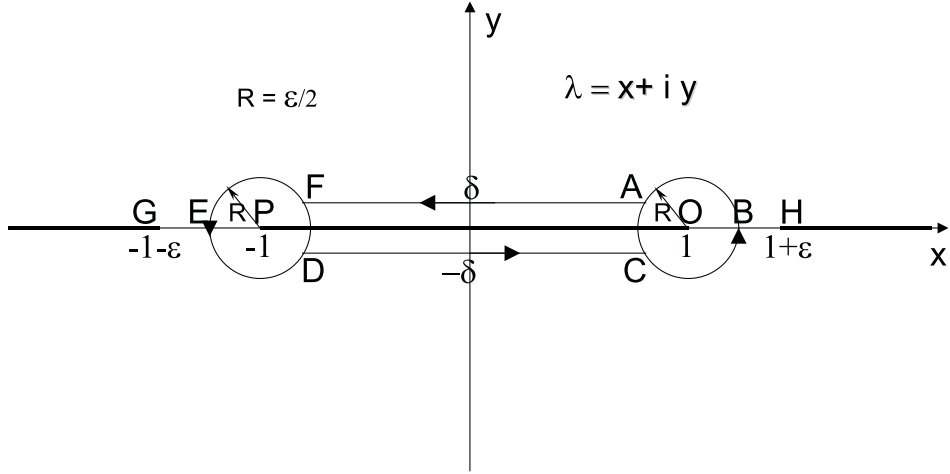}
\end{center}
\caption{\small The contour \mbox{$c(\epsilon,\delta)$}. Bold lines $(-\infty, -1-\epsilon)$ and $(1+\epsilon,\infty)$
are the cuts of integrand $e(1+\epsilon,\lambda)$. Zeros of $D_{\mathrm{L}}(\lambda)$ (Eq.~$\ref{exd}$) are
located on bold line $(-1, 1)$ and this line becomes the cut of $\mathrm{d} \log D_{\mathrm{L}}(\lambda)$ for $L\to \infty$ (Eq.~$\ref{apd}$).
The arrow is the direction of the route of integral we take and $\mathrm{R}$ is the radius of circles.}
\label{fig1}
\end{figure}

However, to obtain all eigenvalues $\nu_m$ directly from matrix
$\mathbf{G}_{\mathrm{L}}$ is a non-trivial task. Let us introduce
\begin{eqnarray}
D_{\mathrm{L}}(\lambda)= \det
(\widetilde{\mathbf{G}}_{\mathrm{L}}(\lambda) \equiv \lambda
I_{\mathrm{L}}- \mathbf{G}_{\mathrm{L}})\;.
\end{eqnarray}
Here $\widetilde{\mathbf{G}}_{\mathrm{L}}$ is a Toeplitz matrix
(see \cite{abbs}) and $I_{\mathrm{L}}$ is the identity matrix of
dimension $\mathrm{L}$. Obviously we also have
\begin{eqnarray}
D_{\mathrm{L}}(\lambda)=\prod_{m=1}^{\mathrm{L}} (\lambda-\nu_m).
\label{exd}
\end{eqnarray}
From the Cauchy residue theorem and analytical property of
\mbox{$e(x, \nu)$}, then $S(\rho_A)$ can be rewritten as
\begin{eqnarray}
S(\rho_A)=\lim_{\epsilon \to 0^+} \lim_{\delta \to
0^+}\frac{1}{2\pi \mathrm{i}} \oint_{c(\epsilon,\delta)}
e(1+\epsilon, \lambda) \mathrm{d} \ln
D_{\mathrm{L}}(\lambda)\;.\label{eaa}
\end{eqnarray}
Here the contour \mbox{$c(\epsilon,\delta)$} in Fig~$\ref{fig1}$ encircles all zeros of \mbox{$D_{\mathrm{L}}(\lambda)$},
but function \mbox{$e(1+\epsilon, \lambda)$} is analytic within the contour.
Just like Toeplitz matrix $\mathbf{G}_{\mathrm{L}}$ is generated
by function $g(\theta)$ in Eqs.~$\ref{mg1}$ and $\ref{mg2}$ [see
next section], Toeplitz matrix
$\widetilde{\mathbf{G}}_{\mathrm{L}}(\lambda)$ is generated by
function $\tilde{g}(\theta)$ defined by
\begin{eqnarray}
\tilde{g}(\theta)=\left\{ \begin{array}{rl}
\lambda- 1, & -k_F < \theta <k_F,\\
\lambda+1, & k_F < \theta < (2\pi-k_F).
\end{array} \right.\label{xxsym}
\end{eqnarray}
Notice that \mbox{$\tilde{g}(\theta)$} is a piecewise constant
function of $\theta$ on the unit circle, with jumps at
\mbox{$\theta=\pm k_F$}. Hence, if one can obtain the determinant
of this Toeplitz matrix analytically, one will be able to get a
closed analytical result for $S(\rho_A)$ which is our new result.
Now, the calculation of $S(\rho_A)$ reduces to the calculation of
the determinant of Toeplitz matrix
$\widetilde{\mathbf{G}}_{\mathrm{L}}(\lambda)$. Before we
calculate the determinant of Toeplitz matrix
$\widetilde{\mathbf{G}}_{\mathrm{L}}(\lambda)$, we also want to
point out two special cases which allow us to obtain an explicit
form for these eigenvalues $\nu_m$ and hence the entropy
$S(\rho_A)$. These are cases with small lattice size of subsystem
A and magnetic $h$ close to critical values $\pm 2$. Let us take
$\tilde{k}_F= k_F$ for $k_F < \frac{\pi}{2}$ or $\tilde{k}_F= \pi-
k_F$ for $k_F > \frac{\pi}{2}$. For cases $\tilde{k}_F \, L \ll
1$, Toeplitz matrix $\mathbf{G}_{\mathrm{L}}$ can be well
approximated by a matrix with diagonal elements
\mbox{($2\tilde{k}_F/\pi-1$)} and all other matrix elements equal
to \mbox{$2\tilde{k}_F/\pi$}. Hence, we can obtain all eigenvalues
for Toeplitz matrix $\mathbf{G}_{\mathrm{L}}$ as $\{ {2
\mathrm{L}\tilde{k}_F/\pi -1}, -1, -1, \cdots, -1\}$  and
$S(\rho_A)$
 becomes
\begin{eqnarray}
S(\rho_A)\approx\frac{2\mathrm{L}\tilde{k}_F}{\pi} \ln
\frac{\pi}{2\mathrm{L}\tilde{k}_F},~~0<\tilde{k}_F \mathrm{L} \ll
1. \label{enf1}
\end{eqnarray}
Expression above can also be re-expressed in terms of $h$ as
\begin{eqnarray}
S(\rho_A)\approx \frac{2\mathrm{L} (1-h^2/4)^{\frac{1}{2}}}{\pi}
\ln \frac{\pi}{2\mathrm{L}
(1-h^2/4)^{\frac{1}{2}}},~~0<(1-h^2/4)^{\frac{1}{2}} \mathrm{L}
\ll 1. \label{enf}
\end{eqnarray}

\section{Toeplitz Matrix and Fisher-Hartwig Conjecture}

Toeplitz matrix \mbox{$T_{\mathrm{L}}[\phi]$} is said to be
generated by function \mbox{$\phi(\theta)$} if
\begin{eqnarray}
T_{\mathrm{L}}[\phi]= (\phi_{i-j}),~~~i,j=1,\cdots,\mathrm{L}-1
\end{eqnarray}
where
\begin{eqnarray}
\phi_l=\frac{1}{2\pi} \int_0^{2\pi} \phi (\theta) e^{-\mathrm{i} l \theta} \mathrm{d} \theta
\end{eqnarray}
is the $l$-th Fourier coefficient of generating function
\mbox{$\phi(\theta)$}. The determinant of
\mbox{$T_{\mathrm{L}}[\phi]$} is denoted by $D_{\mathrm{L}}$. The
asymptotic behavior of the determinant of Toeplitz matrix with
singular generating function was initiated by T.T.\ Wu \cite{wu}
in his study of spin correlation in two-dimensional Ising model
and later incorporated into a more general conjecture, i.e., the
famous Fisher-Hartwig conjecture \cite{fisher,basor,tracy,abbs}.

\noindent {\bf Fisher-Hartwig Conjecture:} Suppose the generating
function of Toeplitz matrix $\phi (\theta)$ is singular in the
following form
\begin{equation}
\phi(\theta)=\psi(\theta) \prod_{r=1}^R t_{\beta_r,\,
\theta_r}(\theta) u_{\alpha_r,\,\theta_r}(\theta)
\end{equation}
where
\begin{eqnarray}
t_{\beta_r,\,\theta_r}(\theta)&=&\exp [-i\beta_r (\pi-\theta+\theta_r)], \qquad \theta_r<\theta <2\pi+\theta_r\label{tdf}\\
u_{\alpha_r,\,\theta_r}(\theta)&=&\Bigl(2-2\cos
(\theta-\theta_r)\Bigr)^{\alpha_r},\quad\quad \Re
\alpha_r>-\frac{1}{2}\label{udf}
\end{eqnarray}
and $\psi$: $\mathbf{T}\to \mathbf{C}$ is a smooth non-vanishing
function with zero winding number. Then as $n\to \infty$, the
determinant of $T_{\mathrm{L}}[\phi]$
\begin{eqnarray}
D_{\mathrm{L}}= \left({\cal F}[\psi]\right)^{\mathrm{L}}
\left(\prod_{i=1}^R {\mathrm{L}}^{\alpha_i^2-\beta_i^2}\right)
{\cal E}[\psi, \{\alpha_i\}, \{\beta_i\},\{\theta_i\}],
~~\mathrm{L}\to \infty.\label{fh}
\end{eqnarray}
Here ${\cal F}[\psi]=\exp \left(\frac{1}{2\pi} \int_0^{2\pi}\ln
\psi(\theta) \mathrm{d} \theta\right)$.
 Further assuming that there exists Weiner-Hopf factorization
\begin{equation}
\psi(\theta)= {\cal F}[\psi]\,
\psi_+\Bigl(\exp(\mathrm{i}\theta)\Bigr)\,
\psi_-\Bigl(\exp(-\mathrm{i}\theta)\Bigr),
\end{equation}
then constant \mbox{${\cal E}[\psi, \{\alpha_i\},
\{\beta_i\},\{\theta_i\}]$} in Eq.~\ref{fh} can be written as
\begin{eqnarray}
{\cal E}[\psi, \{\alpha_i\},\{\beta_i\},\{\theta_i\}]&=&{\cal E}[\psi]
\prod_{i=1}^R G(1+\alpha_i+\beta_i) G(1+\alpha_i-\beta_i)/G(1+2\alpha_i) \nonumber\\
&\times &\prod_{i=1}^R \biggl(\psi_-\Bigl(\exp(\mathrm{i}
\theta_i)\Bigr)\biggr)^{-\alpha_i-\beta_i} \biggl(\psi_+
\Bigl(\exp(- \mathrm{i} \theta_i)\Bigr)\biggr)^{-\alpha_i+\beta_i}\nonumber\\
&\times& \prod_{1\leq i \neq j \leq R}
\biggl(1-\exp\Bigl(\mathrm{i}
(\theta_i-\theta_j)\Bigr)\biggr)^{-(\alpha_i+\beta_i)(\alpha_j-
\beta_j)},
\end{eqnarray}
$G$ is the Barnes $G$-function, \mbox{${\cal
E}[\psi]=\exp(\sum_{k=1}^{\infty} k s_k s_{-k})$}, and $s_k$ is
the $k$-th Fourier coefficient of \mbox{$\ln \psi(\theta) $}. The
Barnes $G$-function is defined as
\begin{eqnarray}
G(1+z)=(2\pi)^{z/2} e^{-(z+1)z/2-\gamma_E z^2/2}
\prod_{n=1}^{\infty} \{ (1+z/n)^n e^{-z+z^2/(2n)}\},
\end{eqnarray}
where $\gamma_E$ is Euler constant and its numerical value is
\mbox{$0.5772156649\cdots$}. This conjecture has not been proven
for general case. However, there are various special cases for
which the conjecture was proven.

For our case, the generating function $\tilde{g}(\theta)$ has two
jumps at $\theta=\pm k_F$ and it has the following canonical
factorization
\begin{eqnarray}
\tilde{g}(\theta)= \psi(\theta)
t_{\beta_1(\lambda),\,k_F}(\theta)t_{\beta_2(\lambda),\,
-k_F}(\theta) \label{bbta1}
\end{eqnarray}
 with
\begin{eqnarray}
\psi(\theta)=(\lambda+1)\left(\frac{\lambda+1}{\lambda-1}\right)^{-k_F/\pi},
~~~\beta(\lambda)=
-\beta_1(\lambda)=\beta_2(\lambda)=\frac{1}{2\pi \mathrm{i}} \ln
\frac{\lambda+1}{\lambda-1}. \label{bbta2}
\end{eqnarray}
The function $t$ was defined in Eq.~$\ref{tdf}$. We fix the branch
of the logarithm in the following way
\begin{eqnarray}
 -\pi \leq \arg \left(\frac{\lambda+1}{\lambda-1}\right) < \pi. \label{bbta3}
\end{eqnarray}
For $\lambda \notin [-1,1]$, we know that $|\Re (\beta_1(\lambda))
|<\frac{1}{2}$ and $| \Re (\beta_2(\lambda)) | < \frac{1}{2}$ and
Fisher-Hartwig conjecture was {\bf PROVEN} by E.L.\ Basor for this
case \cite{basor}. Therefore, we will call it the theorem instead
of conjecture for our application.  Hence following the theorem in
Eq.~$\ref{fh}$, the determinant $D_{\mathrm{L}}(\lambda)$ of
$\lambda I_{\mathrm{L}} -\mathbf{G}_{\mathrm{L}}$ can be
asymptotically represented as
\begin{eqnarray}
D_{\mathrm{L}}(\lambda)&=&\Bigl(2-2\cos(2
k_F)\Bigr)^{-\beta^2(\lambda)}
\left\{G\Bigl(1+\beta(\lambda)\Bigr) G\Bigl(1-\beta(\lambda)\Bigr)\right\}^2\nonumber \\
&&\left\{(\lambda+1)\Bigl((\lambda+1)/(\lambda-1)\Bigr)^{-k_F/\pi}\right\}^{\mathrm{L}}
\mathrm{L}^{-2 \beta^2(\lambda)}.\label{apd}
\end{eqnarray}
Here $\mathrm{L}$ is the length of sub-system A and $G$ is the
Barnes $G$-function and
\begin{eqnarray}
G(1+\beta(\lambda))
G(1-\beta(\lambda))=e^{-(1+\gamma_E)\beta^2(\lambda)}\prod_{n=1}^{\infty}
 \left\{\left(1-\frac{\beta^2(\lambda)}{n^2}\right)^n e^{\beta^2(\lambda)/n^2} \right\}.
\end{eqnarray}

\section{Asymptotic Form of The Entropy}

Now, let us come back to the calculation of entropy $S(\rho_A)$.
For later convenience, let us define
\begin{eqnarray}
\Upsilon(\lambda)=\sum_{n=1}^{\infty} \frac{n^{-1} \beta^2(\lambda)}{n^2-\beta^2(\lambda)}. \label{UPSILON}
\end{eqnarray}
Taking logarithmic derivative of  $D_{\mathrm{L}}(\lambda)$
(Eq.~\ref{apd}), we obtain
\begin{eqnarray}
{\displaystyle \frac{\mathrm{d} \ln D_{\mathrm{L}}(\lambda)} {\mathrm{d} \lambda}}
&=&{\displaystyle \left(\frac{1-k_F/\pi}{1+\lambda} - \frac{k_F/\pi}{1-\lambda}\right)} \mathrm{L}\nonumber\\
&-&{\displaystyle \frac{4}{\mathrm{i} \pi} \frac{\beta(\lambda)
}{(1+\lambda)(1-\lambda)}} \Bigl(\ln \mathrm{L} + \ln (2 |\sin
k_F|)+ (1+\gamma_E) + \Upsilon(\lambda) \Bigr). \label{lnf}
\end{eqnarray}
Let us substitute the asymptotic form above for $\mathrm{d}\ln
D_{\mathrm{L}}(\lambda)/\mathrm{d} \lambda$ into Eq.~$\ref{eaa}$
for entropy $S(\rho_A)$:
\begin{eqnarray}
&{\displaystyle S(\rho_A)=\lim_{\epsilon \to 0^+} \lim_{\delta \to
0^+} \frac{1}{2\pi \mathrm{i}} \oint_{c(\epsilon,\delta)}
e(1+\epsilon, \lambda) \left(\frac{1-k_F/\pi}{1+\lambda} -
\frac{k_F/\pi}{1-\lambda}\right) }\mathrm{L}+&\nonumber\\
&{\displaystyle \lim_{\epsilon \to 0^+} \lim_{\delta \to 0^+}
\frac{2}{\pi^2} \oint_{c(\epsilon,\delta)} \mathrm{d} \lambda
\frac{e(1+\epsilon, \lambda)
\beta(\lambda)}{(1+\lambda)(1-\lambda)} } \Bigl(\ln \mathrm{L} +
\ln (2 |\sin k_F|) + (1+\gamma_E) +\Upsilon(\lambda) \Bigr),&
\label{en1}
\end{eqnarray}
where the contour is taken as shown in Fig.~\ref{fig1}. The first
integral in Eq.~\ref{en1} can be carried out by using the residue
theorem and the definition of function \mbox{$e(x,\nu)$} in
Eq.~$\ref{intee}$. We found that the linear term in $\mathrm{L}$
for entropy $S(\rho_A)$ vanishes.  The second integral can be
calculated as follows: First, we notice that
\begin{eqnarray}
\oint_{c(\epsilon,\delta)} \mathrm{d} \lambda ~(\cdots) = \left(
\int_{\overrightarrow{\textrm{\small
AF}}}+\int_{\overrightarrow{\textrm{\small FED}}}
+\int_{\overrightarrow{\textrm{\small DC}}}
+\int_{\overrightarrow{\textrm{\small CBA}}}\right)
 \mathrm{d} \lambda ~(\cdots)
\end{eqnarray}
Second, we can show that the contribution of integral from the
circular arcs $\overrightarrow{\textrm{\small FED}}$ and
$\overrightarrow{\textrm{\small CBA}}$ vanishes. Therefore, the
entropy (Eq.~$\ref{en1}$) can be written as
\begin{eqnarray}
S(\rho_A)&=& \lim_{\epsilon \to 0^+} \frac{2}{\pi^2}
\Bigl(\int_{1+\mathrm{i} 0^+}^{-1+\mathrm{i} 0^+}
+\int_{-1+\mathrm{i} 0^-}^{1+\mathrm{i} 0^-} \Bigr) \mathrm{d}
\lambda  \frac{e(1+\epsilon, \lambda)
  \beta(\lambda)}{(1+\lambda)(1-\lambda)}\nonumber\\
&\times&  \Bigl(\ln L + \ln (2 |\sin k_F|) + (1+\gamma_E) +
\Upsilon(\lambda) \Bigr).
\end{eqnarray}
For further simplification, we shall use the fact that
\begin{eqnarray}
\beta(x+\mathrm{i} 0^{\pm})=\frac{1}{2\mathrm{i} \pi} \left(\ln \frac{1+x}{1-x}
\mp \mathrm{i}(\pi-0^+)\right)=-\mathrm{i} W(x) \mp (\frac{1}{2} - 0^+)
\end{eqnarray}
for $x\in (-1,1)$ and
\begin{eqnarray}
W(x)=\frac{1}{2\pi }\ln\frac{1+x}{1-x}. \label{wd}
\end{eqnarray}
We can now write the entropy $S(\rho_A)$ as
\begin{eqnarray}
S(\rho_A)&=& \frac{2}{\pi^2} \int_{-1}^1 \mathrm{d} x\,
\frac{e(1, x)}{1-x^2}
\Bigl(\ln \mathrm{L} +\ln (2 |\sin k_F|)+ (1+\gamma_E)\Bigr)\nonumber\\
&+&\sum_{n=1}^{\infty}\frac{2n^{-1}}{\pi^2} \int_{-1}^1 \mathrm{d} x\,
\frac{e(1, x)}{1-x^2} \left( \frac{(\frac{1}{2}+\mathrm{i} W(x))^3}{n^2-(\frac{1}{2}
+\mathrm{i} W(x))^2}+ \frac{(\frac{1}{2}-\mathrm{i} W(x))^3}{n^2-(\frac{1}{2}
-\mathrm{i} W(x))^2} \right), \label{en2}
\end{eqnarray}
where $e(1,x)$ is defined in Eq.~$\ref{intee}$. This expression
for $S(\rho_A)$ contains two integrals. The first integral can be
done exactly as
\begin{eqnarray}
\frac{2}{\pi^2} \int_{-1}^1 \mathrm{d} x\,  (-\frac{1+x}{2} \ln
\frac{1+x}{2}-\frac{1-x}{2} \ln \frac{1-x}{2}) \frac{1}{1-x^2}
 =\frac{1}{3}.
\end{eqnarray}
The second integral in \mbox{Eq.~$\ref{en2}$} can be written as
\begin{eqnarray}
\Upsilon_0&=&\sum_{n=1}^{\infty}\frac{n^{-1}}{\pi^2} \int_{-1}^1
\mathrm{d} x\,
(-\frac{1}{1-x} \ln \frac{1+x}{2}-\frac{1}{1+x} \ln \frac{1-x}{2})\nonumber\\
&&\qquad \times\left( \frac{(\frac{1}{2}+\mathrm{i} W(x))^3}{n^2-(\frac{1}{2}+\mathrm{i}
W(x))^2}+ \frac{(\frac{1}{2}-\mathrm{i} W(x))^3}{n^2-(\frac{1}{2}-\mathrm{i} W(x))^2} \right), \label{cost2}
\end{eqnarray}
which can be further simplified \cite{fff}. Finally we have that
\begin{eqnarray}
S(\rho_A)=\frac{1}{3} \ln \mathrm{L} +\frac{1}{6} \ln
\left(1-\left(\frac{h}{2}\right)^2\right) +\frac{\ln 2}{3}
+\Upsilon_1, ~~\mathrm{L}\to\infty \label{enl}
\end{eqnarray}
with
\begin{eqnarray}
  \Upsilon_1& = &  - \int_0^\infty \mathrm{d} t \left\{ {e^{-t} \over 3 t}
  + {1 \over  t \sinh^2 (t/2)} - { \cosh (t/2) \over 2 \sinh^3 (t/2)}
  \right\}. \label{up12}
\end{eqnarray}
for $\mathrm{XX}$ model. The leading term of asymptotic of the
entropy $\frac{1}{3} \ln \mathrm{L}$ in Eq.~\ref{enl} was first
obtained based on numerical calculation and a simple conformal
argument in Ref.~\cite{vidal,vidal1} in the context of
entanglement. We also want to mention that a complete conformal
derivation for this entropy was found in Ref.~\cite{vk}. One can
numerically evaluate $\Upsilon_1$ to
 very high accuracy to be $0.4950179\cdots$.
 For zero magnetic field
 ($h=0$) case, the constant term $\Upsilon_1 +\ln 2/3$ for $S(\rho_A)$
is close to but different from $(\pi/3) \ln 2$, which can be found
by taking numerical accuracy to be more than five digits.

\section{Summary}

In this paper, we study the entropy of a block of $\mathrm{L}$
neighboring spins in $\mathrm{XX}$ model with the presence of the
transverse magnetic field. We obtain Eq.~\ref{enf} and
Eq.~\ref{enl} for the von Neumann entropy of a block of
$\mathrm{L}$ neighboring spins in $\mathrm{XX}$ with small
$\mathrm{L}$ and large $\mathrm{L}$ respectively. It's interesting
to note that there is a natural length scale
$\mathrm{L}_h=1/\left(1-\left(\frac{h}{2}\right)^2\right)^{\frac{1}{2}}$
for $|h|< 2$ to incorporating the magnetic field effects. When
$|h|$ increases from less than $2$ into larger than $2$, the
system evolves from critical phase into ferromagnetic phase. The
ferromagnetic phase does not have any entropy. The scale
$\mathrm{L}_h$ shows singular behavior at the critical value of
the magnetic field $h_c=2$. We discovered that one can  introduce the universal scaling variable
$\mathcal{L}=2\mathrm{L}/\mathrm{L}_h$:
$$\mathcal{L} \equiv 2\mathrm{L}
\left(1-\left(\frac{h}{2}\right)^2\right)^{\frac{1}{2}}$$ for
$|h|<2$.  Then we can express the von Neumann entropy of
$\mathrm{L}$ neighboring spins in following simple form:
\begin{eqnarray}
S(\rho_A) = \left\{ \begin{array}{ll} \frac{\mathcal{L}}{\pi} \ln
\frac{\pi}{\mathcal{L} } & \textrm{if $0<\mathcal{L}
< 1$}\\
\frac{1}{3} \ln \mathcal{L} +\Upsilon_1 & \textrm{if
$\mathcal{L}\gg 1$}
\end{array} \right.
\end{eqnarray}
with
\begin{eqnarray}
 \Upsilon_1& = & - \int_0^\infty \mathrm{d} t \left\{ {e^{-t} \over 3 t}
  + {1 \over  t \sinh^2 (t/2)} - { \cosh (t/2) \over 2 \sinh^3 (t/2)}
  \right\}.\nonumber
\end{eqnarray}
For small lattice and magnetic field close to $\pm 2$, we obtain
the result directly.  To obtain the result for large $\mathrm{L}$
asymptotically, we first expressed the entropy in terms of the
determinant of a Toeplitz matrix.  Then we used a special case of
Fisher-Hartwig conjecture \cite{fisher} and this special case was
{\bf PROVEN} in \cite{basor}.

From similar calculation, we also obtain the R\'enyi entropy in
Eq.~\ref{olds} to be
\begin{eqnarray}
S_{\alpha}(\rho_A) = \left\{ \begin{array}{ll} \frac{1}{1-\alpha}
\ln
 \left(\left( \frac{\mathcal{L}}{2 \pi}\right)^{\alpha}+
 \left(1- \frac{\mathcal{L}}{2 \pi}\right)^{\alpha}\right) & \textrm{if $0<\mathcal{L}
< 1$}\\
\frac{1+\alpha^{-1}}{6} \ln \mathcal{L} +\Upsilon_1^{\{\alpha\}} &
\textrm{if $\mathcal{L}\gg 1$}
\end{array} \right.
\end{eqnarray}
Here
\begin{eqnarray}
\Upsilon_1^{\{\alpha\}}&=&-\frac{1}{\pi^2} \int_{-1}^1 \mathrm{d}
x\,
 \frac{s_{\alpha}(x)}{1-x^2}
 \left(\psi\Bigl(\frac{1}{2}-\mathrm{i}W(x)\Bigr)
   +\psi\Bigl(\frac{1}{2}+\mathrm{i} W(x)\Bigr)\right),\label{up11}\\
s_{\alpha}(x)&=&\frac{1}{1-\alpha} \ln
 \left(\left(\frac{1+x}{2}\right)^{\alpha}+
\left(\frac{1-x}{2}\right)^{\alpha}\right), \qquad \alpha \neq 1,\label{ssa}\\
\psi (x) &\equiv& {\mathrm{d} \over \mathrm{d} x} \ln \Gamma (x) =
  - \gamma_E + \sum_{n=0}^\infty \left({1 \over n+1} -  {1 \over
  n+x}\right),
  \label{psidef1}\\
W(x)&=&\frac{1}{2\pi }\ln\frac{1+x}{1-x}
   \end{eqnarray}
with $\gamma_E$ the Euler Constant and $\Gamma(x)$ the well-known
Gamma Function.

\section*{Acknowledgments}

We would like to thank Prof. B.M.\ McCoy, Prof. A.\ Abanov and
 Mr. F.\ Franchini for useful  discussions.
This work was supported by National Science Foundation (USA) under Grant
No. DMR-0073058 and PHY-9988566.

\bigskip
\bigskip

\section*{Appendix: Simplification of Formula}

In this appendix, we show more details for simplification of
$\Upsilon_0$ (\ref{cost2}) in detail. In order to simplify
$\Upsilon_0$, we will use the Function $\psi(x)$, which is defined
as
\begin{equation}
  \psi (x) \equiv {\mathrm{d} \over \mathrm{d} x} \ln \Gamma (x) =
  - \gamma_E + \sum_{n=0}^\infty {1 \over n+1} - \sum_{n=0}^\infty {1 \over n+x}
  \label{psidef}
\end{equation}
with $\gamma_E$ the Euler Constant and $\Gamma(x)$ the
well-known Gamma Function, and the property
\begin{equation}
  \psi (x+1) = \psi (x) + {1 \over x}.
  \label{psiprop}
\end{equation}
Introducing $z ~(\overline{z})~ \equiv {1 \over 2} +(-)~
\mathrm{i}
 W(x)$ and using Eqs.~(\ref{psidef}) and (\ref{psiprop}),
 we obtain
\begin{eqnarray}
&&{\displaystyle \sum_{n=1}^\infty n^{-1} \left( {\left( {1\over2} + \mathrm{i} W(x) \right)^3
  \over n^2 - \left( {1\over2} + \mathrm{i} W(x) \right)^2 } +
  {\left( {1\over2} - \mathrm{i} W(x) \right)^3
  \over n^2 - \left( {1\over2} - \mathrm{i} W(x) \right)^2 } \right)}
\nonumber\\
 &=&\psi(1) - 1
  - {1 \over 2} \psi\Bigl( {1 \over 2} - \mathrm{i} W(x) \Bigr)
  - {1 \over 2} \psi\Bigl( {1 \over 2} + \mathrm{i} W(x) \Bigr)
\end{eqnarray}
by using Eq.~(\ref{psiprop})
and definition for $z$ and $\overline{z}$.
Hence, we obtain
\begin{eqnarray}
  \Upsilon_0
  & = & {1 \over \pi^2 }  \int_{-1}^1 \mathrm{d} x \left(
  -{1 \over 1-x} \ln {1+x \over 2} - {1 \over 1+x} \ln {1-x \over 2} \right) \nonumber \\
  && \qquad \times \left[\psi(1) - 1
  - {1 \over 2} \psi\Bigl( {1 \over 2} - \mathrm{i} W(x) \Bigr)
  - {1 \over 2} \psi\Bigl( {1 \over 2} + \mathrm{i} W(x) \Bigr)
  \right]\nonumber\\
&=&\Upsilon_1-\frac{1+\gamma_E}{3}
\end{eqnarray}
with $\Upsilon_1$ defined as
\begin{eqnarray}
\Upsilon_1
  &=&-{1 \over 2 \pi^2 }  \int_{-1}^1 \mathrm{d} x \left(
  -{1 \over 1-x} \ln {1+x \over 2} - {1 \over 1+x} \ln {1-x \over 2} \right) \nonumber \\
  && \qquad \times \left[  \psi\Bigl( {1 \over 2} - \mathrm{i} W(x) \Bigr)
  + \psi\Bigl( {1 \over 2} + \mathrm{i} W(x) \Bigr)
  \right].
\end{eqnarray}
We now perform a change of variable using $w= {1 \over 2 \pi} \ln {1+x \over 1-x}$:
\begin{eqnarray}
  \Upsilon_1 & = & {-2 \over \pi }  \int_0^\infty \mathrm{d} w \left(
  \ln \left[ 2 \cosh (\pi w) \right] - \pi w \tanh (\pi w) \right) \nonumber \\
  && \qquad \qquad \times \left[
   \psi\Bigl( {1 \over 2} - \mathrm{i} w \Bigr)+
   \psi\Bigl( {1 \over 2} + \mathrm{i} w \Bigr) \right].
\end{eqnarray}
We note that
\begin{equation}
  \ln \left[ 2 \cosh (\pi w) \right] - \pi w \tanh (\pi w) = \left.
  \left( 1 - {\mathrm{d} \over \mathrm{d} \alpha} \right)
  \ln \left( 1 + e^{-2 \pi w \alpha} \right) \right|_{\alpha = 1}.
  \label{trans}
\end{equation}
Hence we can rewrite
\begin{equation}
  \Upsilon_1 = {-2 \mathrm{i} \over \pi }  \int_0^\infty \mathrm{d} w \left(
  \ln \left[ 2 \cosh (\pi w) \right] - \pi w \tanh (\pi w) \right) \cdot
  \left( \mathrm{d} \over \mathrm{d} w \right) \ln { \Gamma \left( {1 \over 2} - \mathrm{i} w \right)
  \over \Gamma \left( {1 \over 2} + \mathrm{i} w \right) }.
  \label{bexp}
\end{equation}
Using the following expression for the Logarithm of the Gamma Function:
\begin{equation}
  \ln \Gamma(z) = \int_0^\infty \left[ z-1 - { 1 - e^{-(z-1)t} \over 1 - e^{-t} }
  \right] {e^{-t} \over t} \: \mathrm{d} t
\end{equation}
which is particularly convenient because we need only the imaginary part of it:
\begin{equation}
  \ln { \Gamma \left( {1 \over 2} - \mathrm{i} w \right)
  \over \Gamma \left( {1 \over 2} + \mathrm{i} w \right) } =
  -\mathrm{i} \int_0^\infty \left[ 2 w e^{-t} - {\sin (wt) \over \sinh (t/2) } \right]
  {\mathrm{d} t \over t}.
  \label{gamma}
\end{equation}
After some elementary but tedious calculation, finally we obtain
\begin{eqnarray}
  \Upsilon_1 =  { -} \int_0^\infty \mathrm{d} t \left\{ {e^{-t} \over 3 t}
  + {1 \over  t \sinh^2 (t/2)} - { \cosh (t/2) \over 2 \sinh^3 (t/2)} \right\}.
\end{eqnarray}


\begin{thebibliography}{99}


\bibitem{ben}{C.H.\ Bennett, H.J.\ Bernstein, S.\ Popescu, and B.\ Schumacher, Phys. Rev. {\bf A 53}, 2046, (1996)}

\bibitem{renyi}{A.\ R\'enyi, {\it Probability Theory}, North-Holland,
    Amsterdam, 1970 }

\bibitem{abe}{S.\ Abe and A.\ K.\ Rajagopal, Phys. Rev. {\bf A 60}, 3461,
  (1999)}

\bibitem{Lieb}{E.\ Lieb, T.\ Schultz and D.\ Mattis, Ann. Phys. {\bf 16}, 407, (1961)}

\bibitem{mccoy}{E.\ Barouch and B.M.\ McCoy, Phys. Rev. {\bf A 3}, 786, (1971)}

\bibitem{mccoy2}{E.\ Barouch, B.M.\ McCoy and M.\ Dresden, Phys. Rev. {\bf A 2}, 1075, (1970)}

\bibitem{gallavotti}{D.B.\ Abraham, E.\ Barouch, G.\ Gallavotti and A.\ Martin-L\"of,
    Phys. Rev. Lett. {\bf 25}, 1449, (1970); Studies in Appl. Math. {\bf 50}, 121, (1971); {\cal ibid} {\bf 51}, 211, (1972)}

\bibitem{tsallis}{C. Tsallis, J. Stat. Phys. {\bf 52}, 479, (1988)}

\bibitem{korepin}{N.M.\ Bogoliubov, A.G.\ Izergin, and V.E.\ Korepin,
    {\it Quantum Inverse Scattering Method and Correlation Functions},
    Cambridge Univ. Press, Cambridge, 1993}

\bibitem{vidal}{G.\ Vidal, J.I.\ Latorre, E.\ Rico, and A.\ Kitaev,
    Phys. Rev. Lett. {\bf 90}, 227902, (2003)}

\bibitem{vidal1}{J.I.\ Latorre, E.\ Rico, and G.\ Vidal,
arXiv: quant-ph/0304098}

\bibitem{Henley}{C.A.\ Cheong and C.L.\ Henley, arXiv:
    cond-mat/0206196}

\bibitem{wu}{T.T.\ Wu, Phys. Rev. {\bf 149}, 380, (1966)}

\bibitem{fisher}{M.E.\ Fisher and R.E.\ Hartwig, Adv. Chem. Phys. {\bf 15}, 333, (1968)}

\bibitem{basor}{E.L.\ Basor, Indiana Math. J. {\bf 28}, 975, (1979)}

\bibitem{tracy}{E.L.\ Basor and C.A.\ Tracy, Physica {\bf A 177}, 167, (1991)}

\bibitem{abbs}{A.\ B\"ottcher and B.\ Silbermann, {\it Analysis of
      Toeplitz Operators}, Springer-Verlag, Berlin, 1990}

\bibitem{fff}{F.\ Franchini, {\it Private communication}}

\bibitem{vk}{V.E.\ Korepin, arXiv: cond-mat/0311056}

\end{thebibliography}
\end{document}